# TiS$_3$ transistors with tailored morphology and electrical properties

*Joshua O. Island\*, Mariam Barawi, Robert Biele, Adrián Almazán , José M. Clamagirand, José R. Ares, Carlos Sánchez, Herre S.J. van der Zant, José V. Álvarez, Roberto D'Agosta, Isabel J. Ferrer\*, and Andres Castellanos-Gomez\**

J. O. Island[1], Prof. H.S.J. van der Zant[1], Dr. A. Castellanos-Gomez[1,2]
[1] Kavli Institute of Nanoscience, Delft University of Technology, Lorentzweg 1, 2628 CJ Delft, The Netherlands.
[2] Instituto Madrileño de Estudios Avanzados en Nanociencia (IMDEA-Nanociencia) 28049 Madrid, Spain
E-mail: j.o.island@tudelft.nl, andres.castellanos@imdea.org

M. Barawi[3], J.M. Clamagirand[3], Prof. J.R. Ares[3], Prof. C. Sánchez[3,4], Prof. I.J. Ferrer[3,4]
[3] Materials of Interest in Renewable Energies Group (MIRE Group), Dpto. de Física de Materiales, Universidad Autónoma de Madrid, UAM, 28049- Madrid, Spain
[4] Inst. Nicolas Cabrera, Univ. Autónoma de Madrid, 28049- Madrid, Spain
E-mail: isabel.j.ferrer@uam.es

R. Biele[5], Prof. R. D'Agosta[5,6]
[5] ETSF Scientific Development Center, Departamento de Fısica de Materiales, Universidad del Paı́s Vasco, E-20018 San Sebastian, Spain
[6] IKERBASQUE, Basque Foundation for Science, E-48013, Bilbao, Spain

A. Almazán[7] Prof. J.V. Alvarez[4,7,8]
[4] Inst. Nicolas Cabrera, Univ. Autónoma de Madrid, 28049- Madrid, Spain
[7] Dpt. Física de la Materia Condensada, Univ. Autónoma de Madrid,28049, Madrid, Spain
[8] IFIMAC, Univ. Autónoma de Madrid, 28049- Madrid, Spain

Two dimensional (2D) materials have established their place as candidates for next generation electronics because they present outstanding mechanical and electrical properties, and also because they can be easily integrated with conventional silicon technologies and fabrication protocols. In comparison with their higher (3D) and lower (1D) dimensional counterparts, 2D materials offer a combination of high electric field tunability and large surface area to fabricate complex devices and sensors. The most well-known members of the 2D family (graphene, transition metal dichalcogenides, black phosphorus) consist of materials with weak out-of-plane bonding between layers that allow exfoliation of the 3D bulk down to 2D atomic layers through simple mechanical exfoliation techniques. [1-9] In the layered, van der Waals family, although significantly less explored, are the quasi-one dimensional, transition metal trichalcogenides (TiS$_3$, TaS$_3$, NbS$_3$, NbSe$_3$).[10-13] In particular, TiS$_3$, having a direct optical band gap of ≈1 eV and ultrahigh optical responsivities (≈3000 A/W), shows promise as a suitable replacement to microstructured silicon in applications where high gain is required. [14, 15] The most common synthesis methods for the trichalcogenides (chemical vapour transport (CVT), sulfurization of bulk metal) typically result in films composed of 1D-whiskers and nanoribbons which hampers exfoliation of the bulk material and ultimately limits their integration with other electronic components in nano-devices.[16, 17] Here, we demonstrate control over the morphology of TiS$_3$ to obtain nanosheets with large surface area that facilitates the exfoliation of TiS$_3$ down to a single layer (≈0.9 nm thick). Through extensive characterization, we show that at low growth temperature (400 ºC) TiS$_3$ displays a 2D morphology of flower-like nanosheets whereas at higher growth temperature (500 ºC) TiS$_3$ grows adopting the commonly observed 1D morphology of belt-like nanoribbons. We isolate and exfoliate the bulk TiS$_3$ material obtained at different growth temperatures and systematically compare the electronic properties of the two morphologies of TiS$_3$ (1D nanoribbons and 2D sheets) by fabricating field-effect transistors (FETs). We find that the TiS$_3$ nanosheets, grown at lower temperature, show high electron mobility (up to 73 cm$^2$/Vs) while the TiS$_3$ nanoribbons, on the other hand, have lower mobilities but show a superior electric-field and optical tunability. We attribute the differences in the electronic properties to a higher density of sulphur vacancies in the TiS$_3$ samples grown at lower temperature which, according to density functional theory (DFT) calculations, leads to an increase in the n-type doping level. This work constitutes a first step towards exploiting the almost unexplored trichalcogenide family in 2D electronics applications such as chemical sensors or van der Waals heterostructures where a large surface area is be highly beneficial.

TiS$_3$ is a quasi-one dimensional semiconductor with an optical band-gap of ≈1 eV. **Figure 1(a)** shows the structure of TiS$_3$ which crystalizes in the monoclinic phase. Parallel chains of triangular prisms (TiS$_3$) make up sheets that are held together by van der Waals forces. The parallel chains are responsible for the quasi-one dimensional nature of the material which leads to an anisotropy in the conductivity between the in-plane *a* and *b* axes.[18] Films of TiS$_X$ material can be obtained by direct reaction of titanium and sulphur in a sealed ampule.[19] In the presence of sulphur excess (>75 atomic % sulphur) and at a growth temperature below 632 °C, titanium-sulphur reactants result in the growth of TiS$_3$.[20] Above the critical temperature of 632 °C TiS$_3$ decomposes to TiS$_2$. TiS$_3$ studied here is obtained from the sulphuration of bulk titanium disks.[21] Titanium disks are vacuum sealed in an ampule with sulphur powder (>75 atomic % sulphur) and heated to a designated growth temperature (400 °C to 500 °C). After 20 hours of growth, the ampule is cooled in ambient conditions. The resulting film is characterized by scanning electron microscopy (SEM), X-ray diffraction (XRD), energy dispersive analysis of X-rays (EDX), ultraviolet visible spectroscopy (UV-Vis), and Raman spectroscopy. Interestingly, the morphology of the obtained TiS$_3$ strongly depends on the growth temperature as seen in Figure 1 which shows SEM images of separate TiS$_3$ films grown at 400 °C (Figure 1(b)), 450 °C (Figure 1(c)), and 500 °C (Figure 1(d)). While the 500 °C film is completely composed of belt-like nanoribbons of varying sizes, the 400 °C growth contains nanosheets (fan-shaped petals). The 450 °C growth results in a combination of the two morphologies. Similar types of morphological change with growth temperature have been observed for TiS$_2$ and other layered materials.[22-26] XRD spectra of the films, shown in Figures 1(e) and 1(f), reveal that they are formed by monoclinic TiS$_3$ as a unique crystalline phase. The narrow peaks show that the film is well crystallized for both the nanosheets grown at 400 °C (Figure 1(e)) and the nanoribbons grown at 500 °C (Figure 1(f)). The most intense peak for the nanoribbon sample, corresponding to the ⟨012⟩ direction, is six times larger than the others and indicates a strong preferred orientation in this direction. The ⟨012⟩ peak for the nanosheets, on the other hand, show less preferential growth in this direction. Lattice parameters (*a, b, c* and *β*), determined from the XRD patterns are listed in **Table 1** along with those previously reported (JCPDS-ICDD 15-0783). EDX, UV-Vis and Raman spectroscopy (see Supporting Information) are in agreement with the XRD and show that the two morphologies (nanosheets and nanoribbons) crystallize in the monoclinic phase of TiS$_3$ and have a direct band gap of ≈ 1 eV.

To exfoliate nanoribbons and nanosheets we use a variation of the standard mechanical exfoliation technique developed to isolate graphene.[27] TiS$_3$ material is picked up directly from the as-grown film using a dry viscoelastic stamp (Gelfilm from Gelpak). The stamp containing the material is then placed on a substrate (heavily doped Si chips with 285 nm of SiO$_2$) and slowly peeled off leaving exfoliated material on the substrate surface. In some cases, another clean stamp is used to try to re-exfoliate the material further and remove large unwanted flakes or ribbons. **Figure 2** shows optical images (with superimposed atomic force microscopy (AFM) topographic line profiles in the inset) of nanoribbons and nanosheets after exfoliation. The narrow widths and reduced rigidity of the nanoribbons makes it difficult to exfoliate them down to single layers. The ribbons often break into smaller pieces before exfoliating further. Figure 2(a) shows the thinnest exfoliated nanoribbon with a thickness of 13 nm (see also Supporting Information of Ref. [28]). Increasing thicknesses for exfoliated nanoribbon samples are shown in Figure 2(b) 23 nm, 2(c) 36 nm, and 2(d) 59 nm. A calculation of the exfoliation potential for TiS$_3$ (see Supporting Information) shows that the interlayer coupling is comparable with other 2D materials and even smaller than that of MoS$_2$ indicating exfoliation down to single layers should be possible and indeed, we found that nanosheets, with their increased surface area, could be exfoliated down to a single monolayer (Fig. 2(e)) (See Supporting Information for electrical characteristics of this monolayer TiS$_3$ flake). Increasing thicknesses for sheets are shown in Figure 2(f) 4 nm, 2(g) 17 nm, and 2(h) 30 nm. Sheets become less transparent with increased thickness.

To further characterize the two morphologies of TiS$_3$, FETs are fabricated to measure electrical response. Au/Ti contacts are patterned and deposited on exfoliated nanoribbons and sheets using e-beam lithography and thin film metal deposition. Measurements are then made in a vacuum probe station at room temperature. The quasi-one dimensional structure of TiS$_3$ results in anisotropy of the conductivity between the *a* and *b* axes.[18] To fairly compare the electrical response of nanoribbons and sheets, measurements should be made along the same axis. Nanoribbons naturally grow along the *b*-axis (high mobility axis) making measurements along the *a*-axis challenging due to their reduced widths. Therefore, for comparison with nanoribbon FETs, nanosheet devices are fabricated with electrodes at 30° intervals to determine the *b*-axis and measure their anisotropy. **Figure 3** shows two fabricated FET devices having the same approximate thickness but different morphology. Figure 3(a) shows a nanosheet device where voltage bias measurements are performed between two opposite electrodes across the sheet. The *b*-axis for this sheet, indicated in Figure 3(a) with a dotted line, has been determined by measuring the transfer characteristics across the device between two opposing electrodes at varying angles. The polar plot in the inset of Figure 3(b) shows the high mobility and a low mobility directions (outer dashed line is 80 cm$^2$/Vs and the inner dashed line is 40 cm$^2$/Vs), corresponding to the *b*- and *a*-axis of the nanosheet. From low temperature measurements (see Supporting Information) we extract a Schottky barrier of

≈130 meV. As was shown for MoS$_2$, this could be improved by changing the contact metal.[29] The resistivity at zero back-gate voltage is 0.09 Ω·cm which is an order of magnitude lower than reported values for TiS$_3$ whiskers (2 Ω·cm).[30, 31] The transfer curves show n-type behaviour with an ON/OFF ratio of 5. From the transfer curve we extract the FET mobility using the following equation:

$$\mu = \frac{L}{WC_iV_b}\frac{\partial I}{\partial V_g} \tag{1}$$

where *L* is the channel length, *W* is the channel width, $C_i$ is the capacitance to the gate electrode, and $V_b$ is the voltage bias. Using Equation 1 and a parallel plate capacitor model to calculate the capacitance to the back-gate, we estimate a two-terminal mobility of 73 cm$^2$/Vs. This is two times higher than the hall mobility (30 cm$^2$/Vs) measured in TiS$_3$ whiskers.[32]

Moving now to the nanoribbons, a representative device having the same relative thickness as the nanosheet is shown in Fig. 3(d). Measurements are made between two adjacent electrodes. Fig. 3(e) shows the I-$V_b$ characteristics for the nanoribbon device in Fig. 3(d). The resistivity at zero back-gate voltage is 56 Ω·cm for this nanoribbon device which is more than two orders of magnitude larger than the sheet device. From the transfer curves for the nanoribbon device (Fig. 3(f)) we measure an ON/OFF ratio of ≈300 (at a bias voltage of 1V), considerably higher than that measured for the sheet device. A COMSOL multiphysics simulation is used to account for fringing effects in the calculation of the capacitance of the nanoribbon to the back-gate. Using Equation 1, we estimate a two-terminal mobility of 0.5 cm$^2$/Vs which is two orders of magnitude smaller than the sheet device.

Exploring further the differences in the electrical properties of FETs made from these two TiS$_3$ morphologies, we plot the mobility vs. the ON/OFF ratio in **Figure 4(a)** for all measured nanoribbon and nanosheet (see the Supporting Information for FET characteristics for a monolayer and multilayer flakes) devices. The scatterplot shows a clear contrast between the nanoribbons grown at 500 °C and the sheets grown at 400 °C. The sheets have a more doped semiconducting behaviour (high mobility, low ON/OFF ratio) whereas the nanoribbons have a more semiconducting behaviour (lower mobility, high ON/OFF ratio). Furthermore, we measure the photoresponse of nanoribbons compared with the sheets. Previous measurements on TiS₃ nanoribbons have shown that they respond well to light across the visible spectrum.[28] In Fig. 4(b) we plot the responsivities vs. excitation wavelength for a representative nanoribbon and nanosheet FET device. The responsivity ($R = I_{ph}/P$), where $I_{ph}$ is the measured photocurrent and P is the laser power, is measured with the same biasing conditions for each device ($V_b$ = 100 mV, $V_g$ = 40 V). Note that the power (P) is scaled by the surface area of the device to that of the laser spot (200 µm diameter). The nanoribbon device has responsivities which are two orders of magnitude larger than the sheet device across the visible spectrum. We attribute the low ON/OFF ratio, high mobility, and reduced photoresponse of the nanosheets grown at 400 ºC to having a larger density of free charge carriers. The presence of more free charge carriers leads to higher electrostatic screening of the TiS3 nanosheets which leads to the reduced ON/OFF ratio in comparison with the nanoribbons.[33-35] Free charge carriers can also be responsible for the increase in mobility as they partially screen charged trap states in the SiO2 which lead to scattering, previously shown in MoS2 FETs, but further studies are necessary to determine the exact origin of increased mobility.[36]

A plausible explanation for these findings is the presence of sulphur vacancies which have been shown to induce a strong n-type doping in dichalcogenides.[37-40] This is supported by controlled sulphur desorption of nanoribbon films (see Supporting Information) and DFT calculations (below). We find that by controllably annealing nanoribbon films to high temperatures (> 300 ºC) in vacuum, the stoichiometric ratio of the film reduces, indicating an increase in sulphur vacancies. These annealed films with a lower stoichiometric ratio display a lower resistance and Seebeck coefficient value in comparison with the as-grown film, indicating a relationship between the presence of sulphur vacancies and the higher density of free-carriers.

In addition, these experimental findings are supported by a first principle investigation of the electronic structure of TiS$_3$. The electronic band structure for the clean TiS$_3$ and the system with a fairly large density of S vacancies (2%) are reported in Figure 4(c) and (d), respectively (electronic band structure for the mono- and bilayer system can be found in Figure S9 of the Supporting Information). From the comparison between Figure 4(c) and (d), it is seen that the presence of the vacancy creates a localised electronic state below the Fermi energy, therefore the S vacancy acts as n-dopant. This state is indeed fully occupied, and its presence below the Fermi energy increases the free carrier (electron) density. On the other hand, the vacancies do not affect the direct gap at the Gamma point that remains about 0.8 eV. Indeed, the localised vacancy state has a fairly small density of states, and therefore it is unlikely detected while measuring the gap optically. This could explain why the band gap from UV-Vis measurements (see Supporting Information) is the same in the two samples. Furthermore, the calculated energy difference between the ground state energy for the clean and the doped TiS$_3$ is about 300 meV indicating that the vacancy will most likely be filled in a S rich atmosphere at a large temperature. Indeed, in such a configuration, a vacancy will reach the surface and rapidly recombine with a S atom from the atmosphere. This could explain why the nanosheets grown at 400 °C are more rich in vacancies and therefore act as a strongly doped semiconductor. At the same time, it can be seen from Figure 4(d) that the vacancy state below the Fermi energy has some dispersion, i.e., the state is not fully localised, although, the effective mass of the electrons in this state is fairly large. The finite dispersion of the vacancy state is probably an artefact of the fairly large density of vacancies we can investigate within our numerical approach. In our analysis, we have considered all the 3 possible configurations with a vacancy (due to symmetry the other 3 are equivalent to those). We have found that only one configuration is electronically active, the other 2 are higher in energy with respect both to the clean system which provides the ground state configuration and to the vacancy that is electronically more active, and do not significantly affect the electronic properties of TiS$_3$.

We demonstrate that the TMDC TiS3 can be synthesized with a 2D morphology instead of the commonly observed belt-like 1D shape by reducing the growing temperature. We additionally show that this 2D morphology can be exfoliated down to single layers (≈0.9 nm thick) which is more favourable for integration in next generation nanoelectronics and to fabricate van der Waals hetereostructures. Interestingly, we find that apart from the morphological difference, the exfoliated nanoribbons and nanosheets also present different electronic properties. FETs fabricated using nanosheets present high mobilities up to 73 cm$^2$/Vs while those fabricated using individual nanoribbons present lower mobilities but higher gate tunability and optical response. We attribute these differences to a higher density of sulphur vacancies in the nanosheets compared with the nanoribbons which, according to DFT calculations, leads to n-type doping. This demonstrated control over the morphology and electrical properties of TiS$_3$ opens the door to other members of the trichalcogenides family, broadening the collection of materials which can be exfoliated down to a single layer and incorporated in next generation electronics.

**Experimental Section**

*Characterisation of TiS$_3$ nanoribbons and sheets.* SEM, XRD, EDX, UV-Vis, Raman spectroscopy, and AFM are employed to characterize the structure and topography of the as-grown TiS$_3$ nanoribbons and nanosheets. XRD patterns were taken in a Panalytical X'pert Pro

X-ray diffractometer with CuK$\alpha$ radiation ($\lambda$=1.5406 Å) in $\vartheta$-2$\vartheta$ configuration. Stoichiometry of films was determined by EDX from Oxford Instruments mod. INCA x-sight using an incident electron beam energy of 10kV and acquisition times of 100 s. Optical spectral response (UV-Vis) of the TiS$_3$ films deposited on quartz have been measured in order to determine the band gap energy. Raman measurements are performed in a Renishaw *in via* system in backscattering configuration ($\lambda$ = 514 nm, 100× objective with NA = 0.95). This configuration has a typical spectral resolution of ~ 1 cm$^{-1}$. To avoid laser-induced modification or ablation of the samples, all spectra were recorded at low power levels of 100 – 500 µW. AFM is used to measure the thickness of the nanoribbons and nanosheets. The AFM (*Digital Instruments D3100 AFM*) is operated in amplitude modulation mode with Silicon cantilevers (spring constant 40 N m$^{-1}$ and tip curvature <10 nm).

*Fabrication of TiS$_3$ field effect transistors(FETs):* Single nanoribbons and nanosheets are isolated from an as-grown film and subsequently transferred to a SiO$_2$/Si substrate with an all-dry transfer method. Thin nanoribbons and nanosheets are then selected by their optical contrast. We use standard e-beam lithography and lift-off procedures to define the metallic contacts (5 nm Ti/50 nm Au, nanoribbons) (5 nm Ti/100 nm Au, nanosheets).

*Characterisation of the FETs*: FET characterisation is performed in a *Lakeshore Cryogenics* probe station at room temperature in vacuum (<10$^{-5}$ mbar). Laser excitation is provided by diode-pumped solid-state lasers operated in continuous wave mode (CNI Lasers). The light is coupled into a multimode optical fibre through a parabolic mirror. At the end of the optical fibre, another identical parabolic mirror collimates the light exiting the fibre. The beam is then directed into the probe station's zoom lens system and subsequently inside the sample space. The beam spot size on the sample has a 200 µm diameter.

*Band structure calculations:* We have calculated the electronic properties of TiS$_3$ using the Quantum Espresso package.[41] By starting from the lattice parameters provided in this paper, we have built a 2x2x2 supercell to reduce the vacancy density as much as possible. We have therefore optimised the atomic positions with a residual force after relaxation of 0.001 atomic units, for both the pristine system and with a vacancy. All the configurations are ionically relaxed using the Broyden–Fletcher–GoldfarbShann's procedure available within the Quantum Espresso package. The exchange–correlation potential is described self-consistently within the generalised gradient approximation throughout the Perdew–Burke–Ernzerhof's functional, and the Trouiller-Martins' norm-conserving pseudo-potential both for Ti and S is used to model valence electron–nuclei interactions. The energy cut-off for the plane wave basis set is put at 30 Ry with a charge density cut-off of 240 Ry. The path for the calculations of the k-points in the band structure follows the nomenclature for the monoclinic structure reported in Ref. [42]

**Acknowledgements**

This work was supported by the European Union (FP7) through the program RODIN, the Dutch organization for Fundamental Research on Matter (FOM), and NWO/OCW. A.C-G. acknowledges financial support through the FP7-Marie Curie Project PIEF-GA-2011-300802 ('STRENGTHNANO'). Authors from MIRE Group acknowledge the support of the Ministry of Economy and Competitiveness (MINECO) for this research (contract MAT2011-22780). They also acknowledge technical support from Mr. F. Moreno. R.B. and R.D'A. acknowledge the financial support of the Grupos Consolidados UPV/EHU del Gobierno Vasco (IT578-13), CONSOLIDER INGENIO 2010: NANOTherm (CSD2010-00044), and the computational facilities of the Donosti International Physics Center where some of the calculations have been performed. A. A. and J.V.A. acknowledge grant number: FIS2012-37549-C05-03.

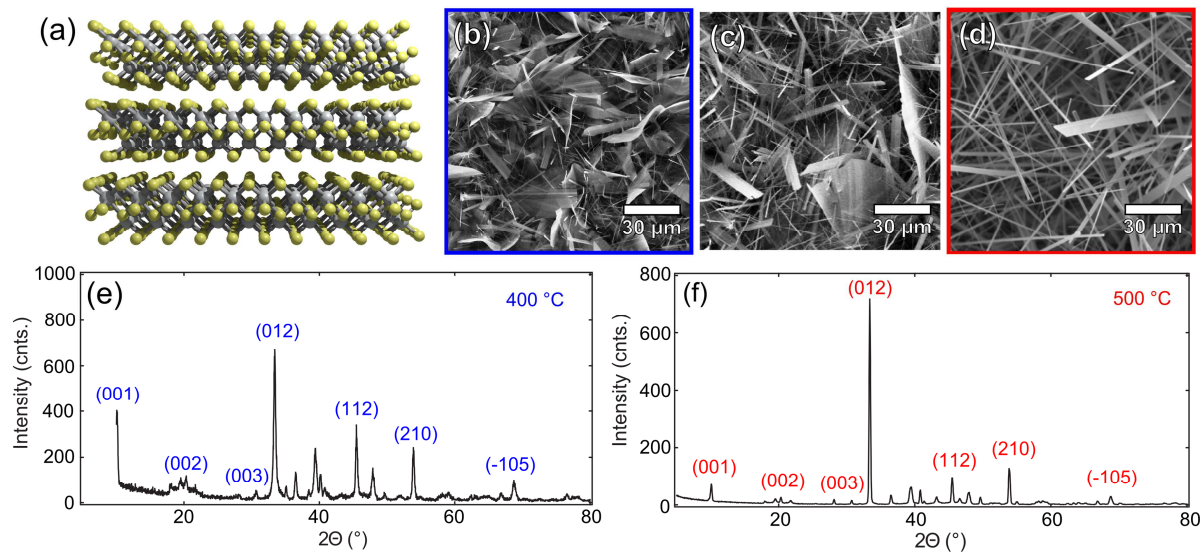

**Figure 1.** (a) Scheme of the TiS₃ crystalline structure. Ti atoms are covalently bonded to six sulphur atoms forming triangular prisms which are stacked to make chains. SEM images of TiS₃ film grown at (b) 400 °C (nanosheets), (c) 450 °C, and (d) 500 °C (nanoribbons). (e) XRD patterns of TiS₃ film formed by nanosheets (400 °C). Some Miller indices for the monoclinic structure (JCPDS-ICDD 15-0783) are included. (f) XRD patterns of TiS₃ film formed by nanoribbons (500 °C). The same Miller indices in (e) are labelled here.

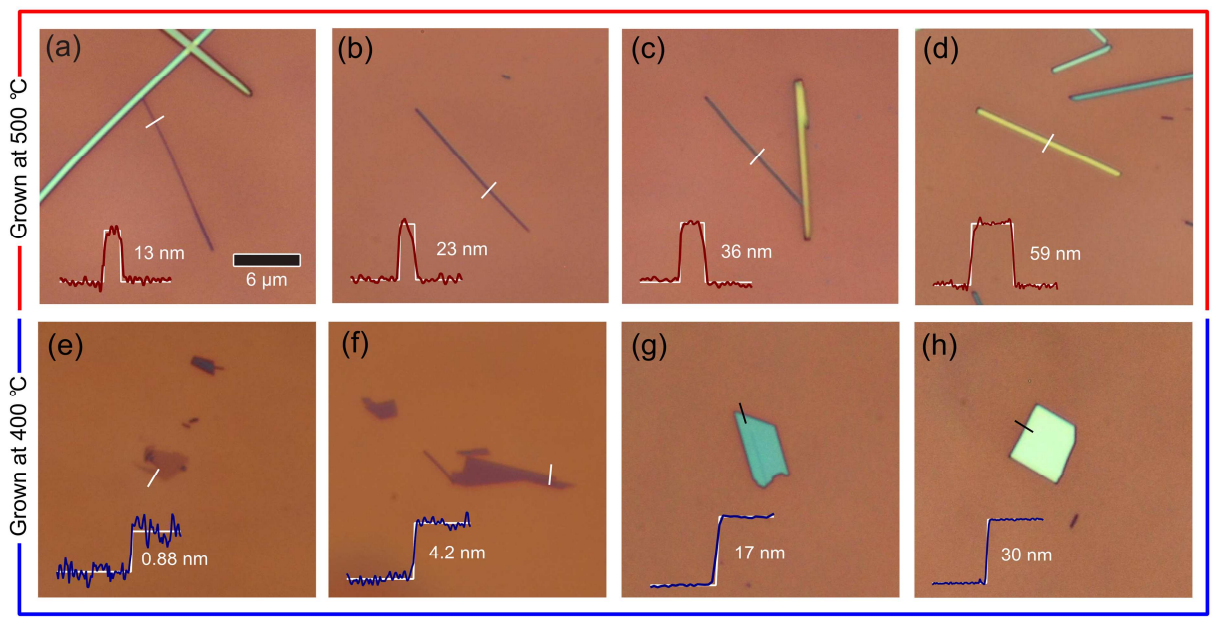

**Figure 2.** Optical images of TiS₃ nanosheets and nanoribbons exfoliated from TiS₃ films sulfurized at 400ºC and 500ºC respectively. The scale bar of 6 µm applies for all panels. (a) Exfoliated 13 nm nanoribbon, (b) 22 nm nanoribbon, (c) 36 nm nanoribbon, and (d) 59 nm

nanoribbon. (e) Exfoliated single-layer TiS$_3$ sheet, (f) 4 nm sheet, (g) 17 nm sheet, and (h) 30 nm sheet.

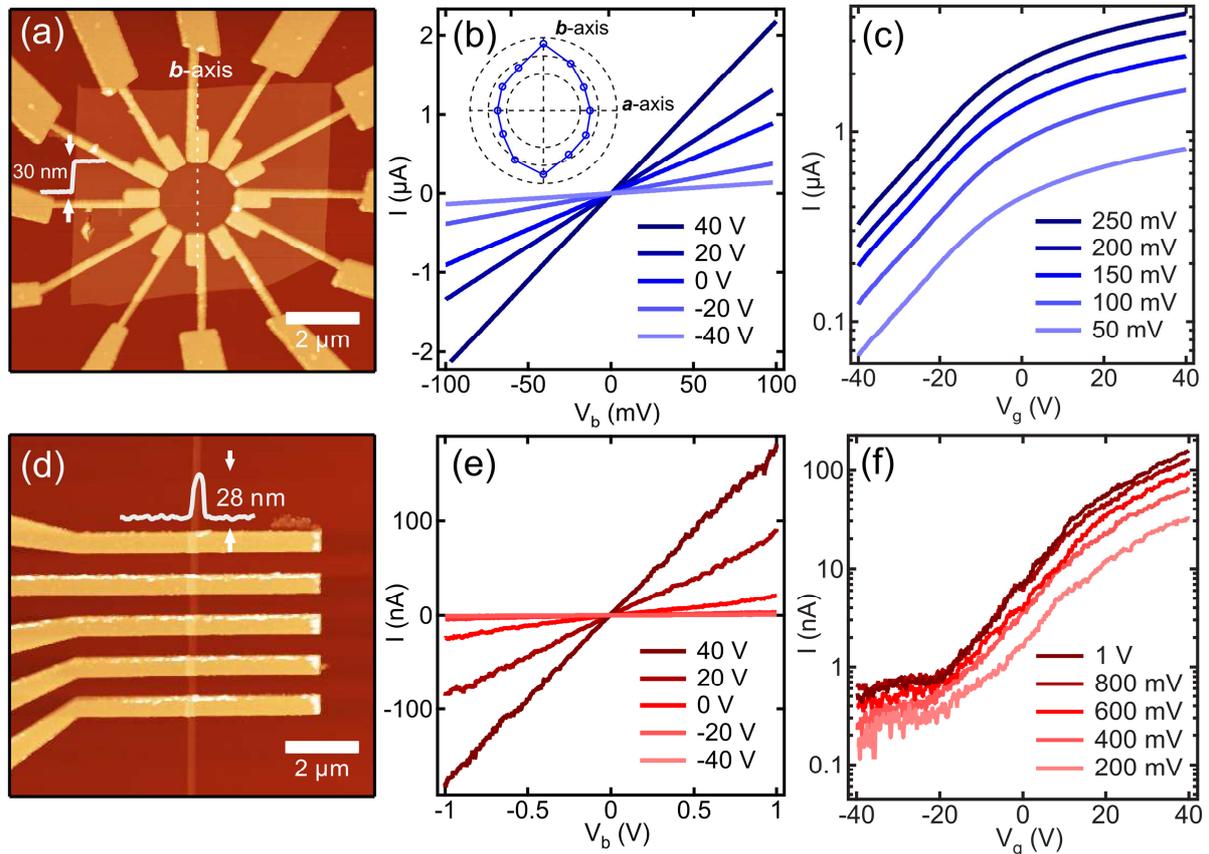

**Figure 3.** Electrical characterization of TiS$_3$ nanosheets and nanoribbons. (a) AFM image of 30 nm sheet. (b) Current-voltage (I-V$_b$) characteristics for sheet device in (a). Inset shows the anisotropy of the mobility at V$_g$ = -10V and V$_b$ = 100mV. The outer dashed line marks a mobility of 80 cm$^2$/Vs and the inner dashed line marks 40 cm$^2$/Vs. The **b**-axis and **a**-axis correspond to the electrode geometry and axes labelled in (a). (c) Transfer (I-V$_g$) characteristics for nanosheet device in (a). (d) AFM of 28 nm nanoribbon. (e) Current-voltage (I-V$_b$) characteristics for the nanoribbon device in (d). (f) Transfer (I-V$_g$) characteristics for the same nanoribbon device.

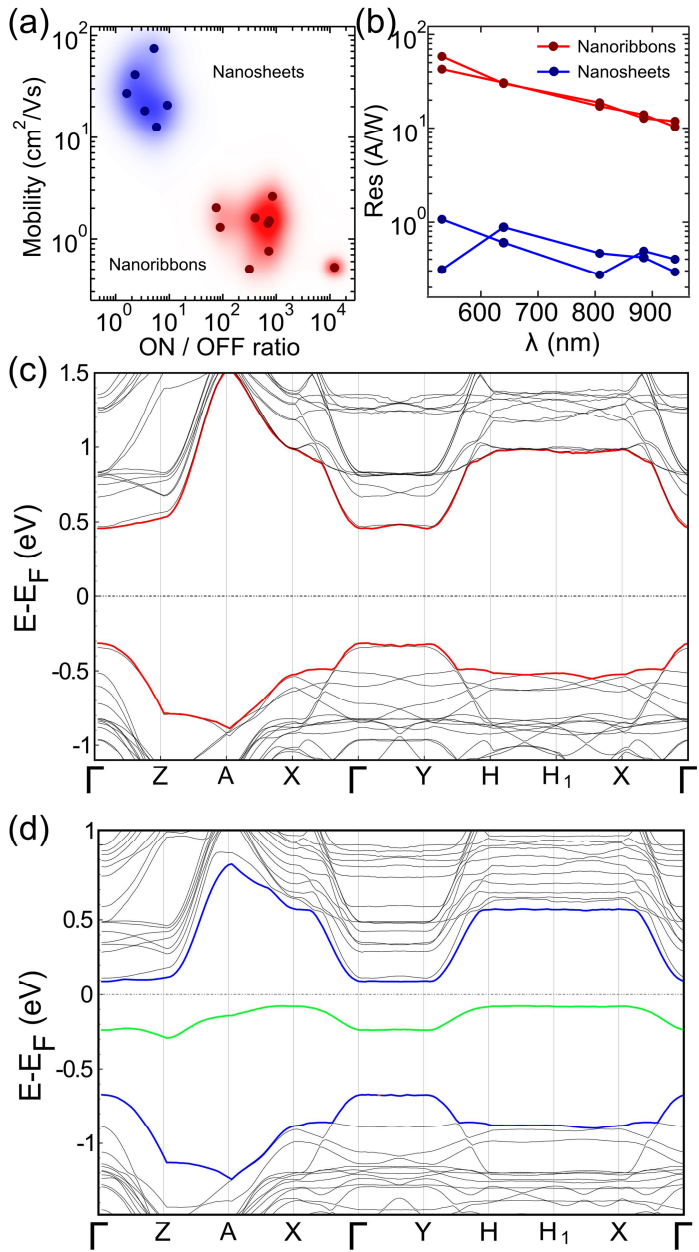

**Figure 4.** (a) Mobility vs. ON/OFF ratio for nanoribbons and sheets. (b) Responsivity vs. wavelength for two representative devices for both nanoribbons and nanosheets using a 500 μW laser excitation ($V_b$ = 100 mV, $V_g$ = 40 V for both devices). (c) Electronic band structure for the clean TiS$_3$ system. (d) Electronic band structure for the doped TiS$_3$ with sulphur vacancies.

**Table 1.** Lattice parameters of TiS$_3$ films grown at 400 °C and 500 °C obtained from XRD patterns along with the values previously reported in the Power diffraction file alphabetical index JCPDS-ICDD 15-0783.

|  | JCPDS | 500 °C | 400 °C |
|---|---|---|---|
| *a* (Å) | 4.973(0) | 4.947(5) | 4.9280 |
| *b* (Å) | 3.433(0) | 3.397(5) | 3.4115 |
| *c* (Å) | 8.714(0) | 8.780(5) | 8.7949 |
| *β* (º) | 97.74(0) | 95.45(5) | 95.27 |
| *V* (Å$^3$) | 147.41(0) | 146.88 (3) | 147.27 |

Supporting Information for:

# TiS$_3$ transistors with tailored morphology and electrical properties

*Joshua O. Island\*, Mariam Barawi, Robert Biele, Adrián Almazán , José M. Clamagirand, José R. Ares, Carlos Sánchez, Herre S.J. van der Zant, J.V. Alvarez, Roberto D'Agosta, Isabel J. Ferrer\*, and Andres Castellanos-Gomez\**

**Supporting Information Contents**
1. **EDX**
2. **UV-Vis**
    **Figure S1:** Band gap energy estimated from UV-Vis measurements
3. **Raman spectra for TiS$_3$ nanoribbons and nanosheets after exfoliation**
    **Figure S2:** Raman spectra for exfoliated nanoribbons and nanosheets
4. **Monolayer and multilayer TiS$_3$ FET characteristics**
    **Figure S3:** FET characteristics of monolayer TiS$_3$ device
    **Figure S4:** FET characteristics for two multilayer devices
5. **Calculated exfoliation potential for TiS$_3$**
    **Figure S5:** Energy vs. interlayer distance
    **Table S1:** Comparison of interlayer potential parameters for different 2D materials
6. **Schottky barrier for TiS3 with Au/Ti contacts**
    **Figure S6:** Schottky barrier estimate from low temperature measurements
7. **Controlled sulphur desorption of a TiS$_3$ film**
    **Figure S7:** XRD spetra before and after sulphur desorption
    **Table S2:** Stoichiometry ratio, Seebeck coefficient, and resistivity before and after sulphur desorption
    **Figure S8:** Seebeck coefficient before and after sulphur desorption
8. **Mono- and bilayer electronic band structure**
    **Figure S9:** Electronic band structure for mono- and bilayer TiS$_3$

1. **EDX**

Stoichiometry of films was determined by an Energy Dispersive x-ray analyzer from Oxford Instruments mod. INCA x-sight using an incident electron beam energy of 10kV and acquisition times of 100 s. Values obtained were, S/Ti=2.89 and S/Ti=2.94 (±0.1) for the nanosheets grown at 400ºC and the nanoribbons grown at 500ºC, respectively.

2. **UV-Vis**

The band gap energy was experimentally investigated in TiS$_3$ thin films by optical spectroscopy[1] obtaining two direct optical energy band gaps at 0.96 ± 0.05 eV and 1.45 eV ±0.05 eV, independently on the temperature and sulfuration time. Minimum band gap energy shows a slight raising trend (red dotted line) with an increase in sulfuration temperature but their values are in the error range of the measurement (Figure S1). Films grown at 500ºC were not measured.

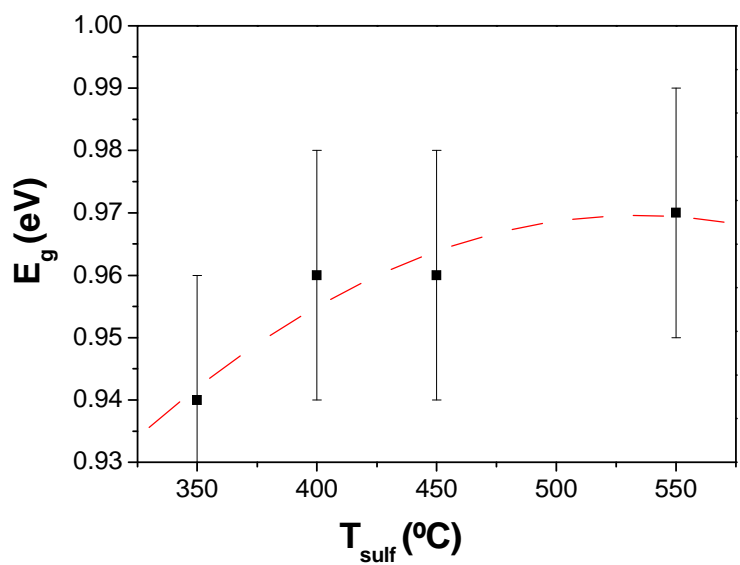

**Figure S1**: Band gap estimation from UV-Vis measurements for different film growth temperatures.

3. **Raman spectra of TiS$_3$ nanoribbons and nanosheets after exfoliation**

Raman spectra were taken on selected nanoribbons and platelets after transfer and exfoliation to confirm their composition (see Figure S2). The spectra reveal three peaks corresponding to the three A$_g$-type modes of TiS$_3$.[2]

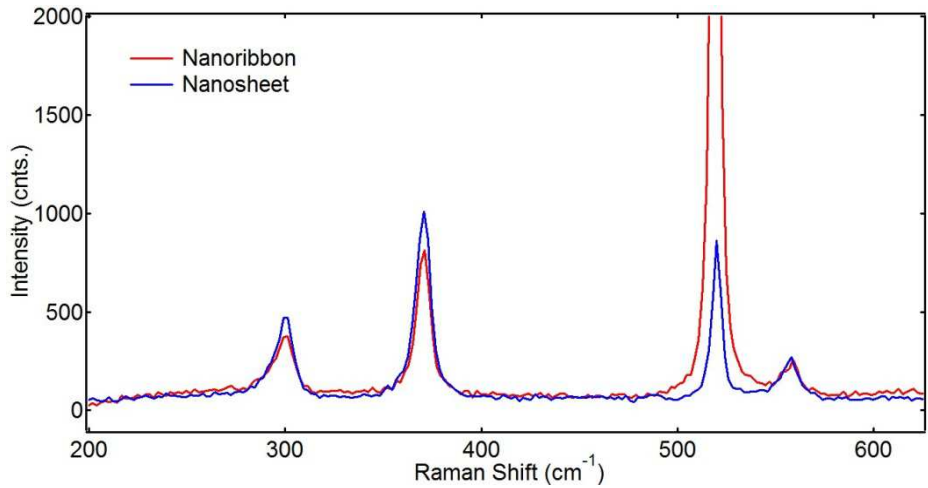

**Figure S2: Raman spectra for a representative nanoribbon (red) and nanosheet (blue) showing three peaks at 300, 370, and 557 cm$^{-1}$ corresponding to A$_g$-type modes. 520 cm$^{-1}$ is the silicon substrate peak.**

4. **Monolayer and multilayer TiS$_3$ FET characteristics**

Using the single layer TiS$_3$ flake presented in the main text (Figure 2(e)), a FET device is fabricated and shown in Figure S3(b). Two terminal measurements are made between the inner two electrodes. Unlike the thicker nanosheet devices, the monolayer device has nonlinear IVs (Figure S3(c)) indicating poor contact to the flake. From the transfer curves in Figure S3(d) we estimate a two-terminal mobility of 2.6 cm$^2$/Vs which is most likely limited by the poor electrode contact.

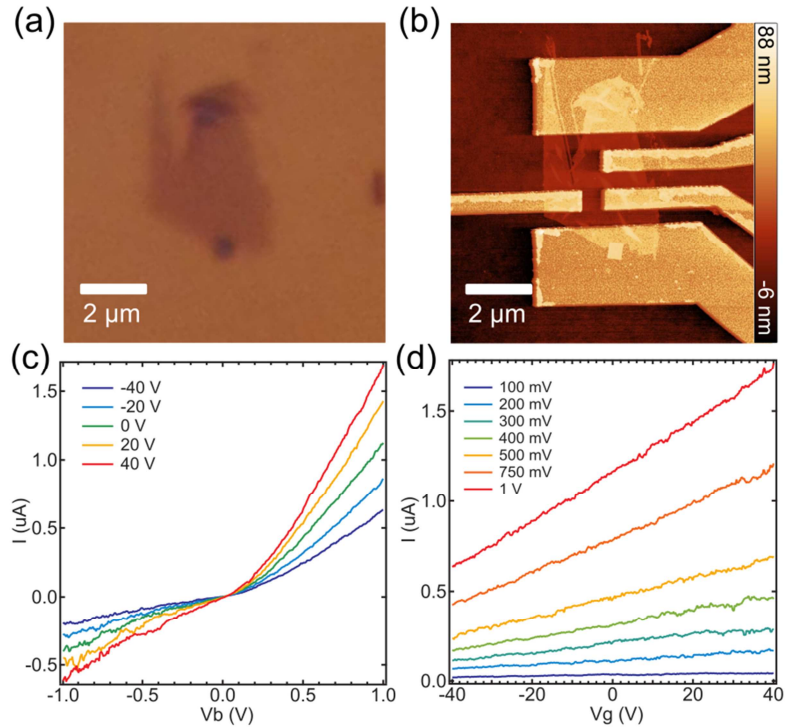

**Figure S3**: Optical image of the monolayer flake before fabricating electrodes. (b) AFM image of the flake in (a) after fabricating electrodes. Note that the colour scale is nonlinear to emphasize the flake. (c) I-$V_b$ characteristics of the FET at different gate voltages. (d) Transfer characteristics of the FET at different bias voltages.

Two multilayer devices are show in Figure S4. These devices, although significantly thinner (4.7 nm and 6.4 nm) than the main text nanosheet device, show comparable FET characteristics. In general, this is consistent with ab inito calculations for a monolayer and bilayer system (see section 8 below) which show little change in the band gap. The multilayer devices have comparable ON/OFF ratios of 4 and 6 and mobilities of 23 cm$^2$/Vs and 24 cm$^2$/Vs for device 1 and 2, respectively.

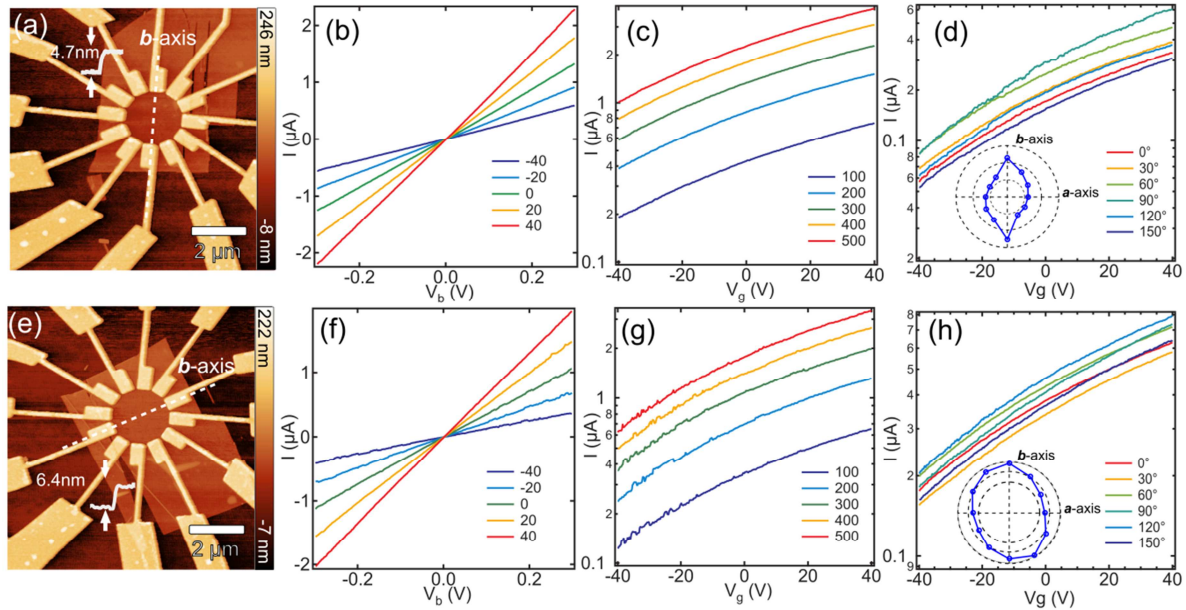

**Figure S4**: (a) AFM image of a 4.7 nm thick nanosheet device. The b-axis is indicated by the white dotted line. Note that the colour scale is nonlinear to emphasize the flake. (b) I-$V_b$ characteristics of the device in (a) along the b-axis. (c) Transfer characteristics at a bias voltage of 100 mV for the device in (a) along the b-axis. (d) Transfer curves at $V_b$=100mV taken at increasing angles relative to the b-axis. The estimated two terminal mobility is given in the inset. The dotted lines mark mobilities of 10, 20, and 30 cm$^2$/Vs. (e) AFM image of a 6.4 nm thick nanosheet device. The b-axis is indicated by the white dotted line. Note that the colour scale is nonlinear to emphasize the flake. (f) I-$V_b$ characteristics of the device in (e) along the b-axis. (g) Transfer characteristics at a bias voltage of 100 mV for the device in (e) along the b-axis. (h) Transfer curves at $V_b$=100mV taken at increasing angles relative to the b-axis. The estimated two terminal mobility is given in the inset. The dotted lines mark mobilities of 5, 15, and 25 cm$^2$/Vs.

## 5. Calculated exfoliation potential for TiS3

To obtain the exfoliation potential we have computed the total energy in a bilayer system at different interlayer distances. The energies were computed with SIESTA[3] using the DRSLL van der Waals density functional of Dion et al.[4]. The density functional is implemented using kernel factorization by Gomez-Román and Soler[5]. We used a basis of DZP numerical atomic orbitals and Troullier-Martins pseudopotentials. The relaxation takes as a starting point the crystal structure of the nanoribbons grown at 500 ºC.

The distance between layers is defined to have direct comparison between the equilibrium point and the lattice parameter c. The results are presented in Figure S5.

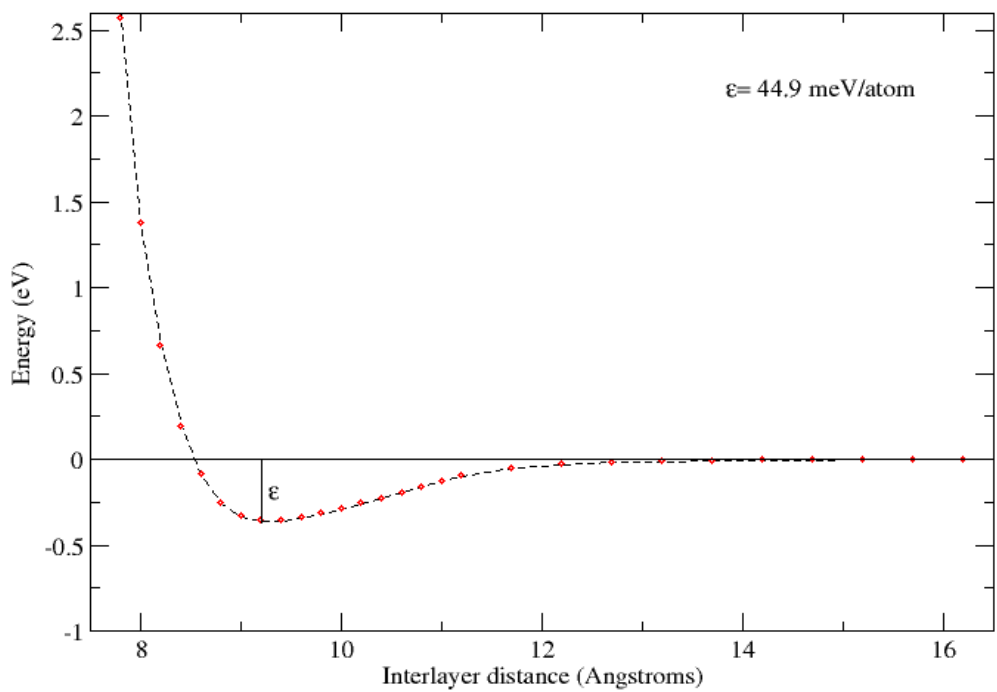

**Figure S5:** Exfoliation potential for TiS$_3$. The interlayer distance is defined to compare with the bulk lattice parameter c. The energy is given in eV but the energy $\varepsilon$ is expressed in meV per atom in the unit cell to compare with other 2D materials. The dashed lines are guides for the eye.

The energy shows the typical binding potential shape. The potential well depth is $\varepsilon = 44.9$ meV/atom and the interlayer distance at the minimum is $d_{min}=9.208$ Å. The value of the interlayer distance of the bilayers at the minimum is larger than the experimental value of c=8.780Å in bulk by a 4.5%. This difference is very similar to the ones found in other 2D materials. For instance in Black phosphorus the difference is 6.2% and in MoS$_2$ it is 4.2%.

In Table S1 we present a comparison with other 2D materials. The $\varepsilon$ parameter in TiS3 has an intermediate value compared to those in other 2D crystals. It has a slightly higher value than phosphorene[6] and graphite[7], but smaller than MoS$_2$[7].

**Table S1**: Comparison of interlayer potential parameters for different 2D materials.

| Material | $\varepsilon$ ( meV/atom) | $d_{min}$ (Å) |
|---|---|---|
| TiS$_3$ | 45 | 9.208 |
| Black Phosphorus[6] | 15 | 5.586 |
| Graphite[7] | 24 | 7.497 |
| MoS$_2$[7] | 60 | 6.302 |

## 6. Schottky barrier for TiS3 with Au/Ti contacts

We extract the *true* Schottky barrier (SB) height by determining the flat band voltage following Das et al.[8] Figure S6(a) shows the low temperature transfer curves for the device shown in Figure 3(a) of the main text. The current through the device from thermionic emission theory is given by:[9]

$$I_d \sim T^2 \exp\left(-\frac{q\Phi_B}{k_b T}\right)$$

where $T$ is the temperature, $q$ is the electronic charge, $k_b$ is the Boltzmann constant, and $\Phi_B$ is the SB. Extraction of the SB is made by linearly fitting the temperature data plotted as ln(I/T^2) vs 1/T (see Figure S6(b)). The slope (S) of the fit gives the SB height, S = (-q$\Phi_B$/$k_b$). The extracted SB height for the full range of gate voltages is plotted in Figure S6(c). When the SB deviates from a linear trend at negative gate voltages, band bending starts to occur and tunnelling current adds to thermionic emission current giving an underestimate of the SB height. The true height is extracted at the flat band voltage (Vg ≈ 0V) and given here as SB ≈ 130 meV. A low bias voltage of 100 mV is used which we expect leads to a difference of 5% in our extracted SB height.[10]

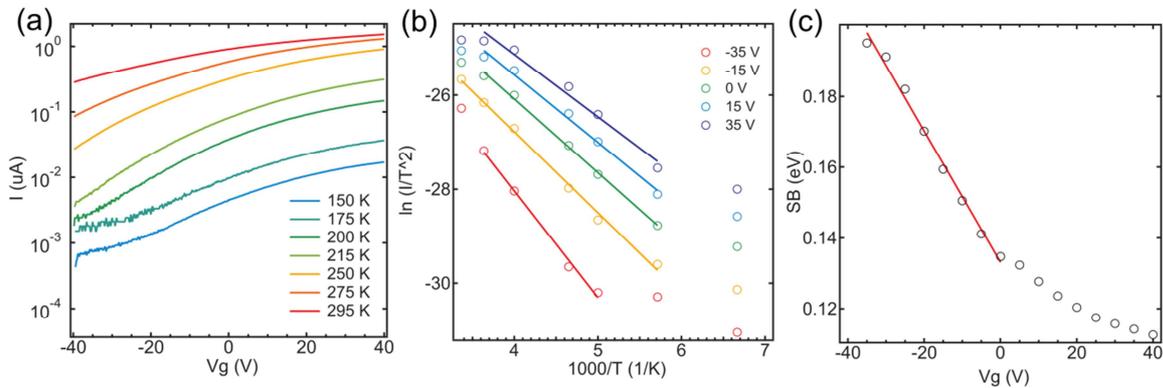

**Figure S6:** (a) Low temperature transfer curves for the device shown in Fig. 3(a) of the main text. (b) Fits to the thermionic emission model to extract the schottky barrier height. (c) Schottky barrier as a function of back gate voltage.

## 7. Controlled sulphur desorption of a TiS₃ film

A Ti film of 300 nm of thickness was sulfurated at 550 ºC during 20h, as usual. After sulfuration, the TiS₃ film was flattened in order to perform measurements of the film resistivity (ρ) and Seebeck coefficient (S). The film thickness after being flattened is 1500 ± 200 nm.

For controlled sulphur desorption, the film was annealed up to T = 360 ºC in vacuum (10⁻² mbar). After the sample reached T = 360ºC, it was cooled down to room temperature. XRD before and after the vacuum sulphur desorption treatment is shown in Figure S7. Both diffraction patterns exhibit identical diffraction peaks related to an unique phase of TiS₃, indicating that the thermal treatment hardly affects the crystalline phase, although the relative peak intensities seem to be affected, i.e. preferential orientation could exist.

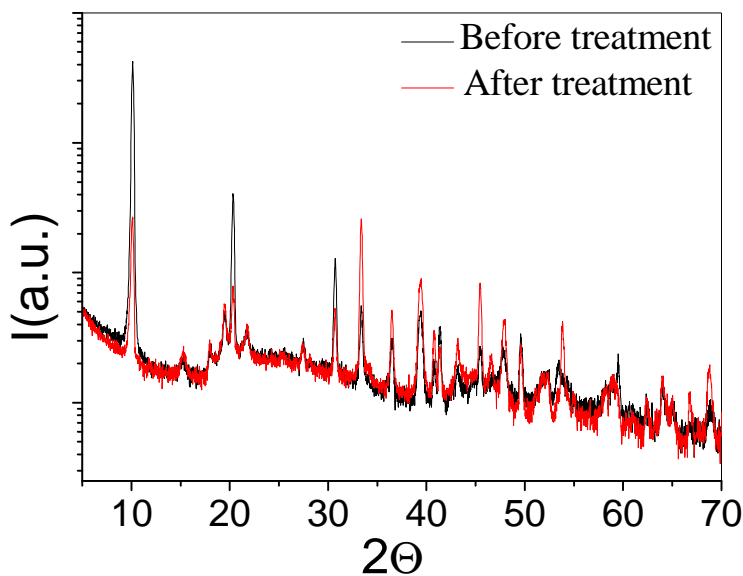

**Figure S7**: XRD pattern diffraction of TiS$_3$ film before and after annealing up to 360ºC.

EDX measurements before and after the treatment reveal, however, a substantial decrease on the sulphide stoichiometry from S/Ti = 3.20 ±0.20 to S/Ti = 2.52 ±0.20. Concerning the transport properties, ρ and S values at RT were reduced from -546 ± 20 µV/K and 0.7± 0.1 Ωcm to -420 ± 20µV/K and 0.40 ±0.05 Ωcm, respectively (see Table S2 and Figure S8). After sulphur desorption the film exhibits a more "metallic" behaviour. The diminution of the stoichiometry could be attributed to sulphur loss from the film due to the high annealing temperature (>300 ºC). Sulphur loss could affect the transport properties (S and ρ) by generating vacancies in the film. In fact, measurements performed by TGA (unpublished) reveal that TiS$_3$ could start to lose sulphur at temperatures near 300ºC depending on the residual atmosphere. In addition, given that the annealing is performed well below the critical decomposition temperature of 632 ºC, the reduction in stoichiometry is most likely due to a significant increase in sulphur vacancies and not a compositional change to TiS$_2$.[11]

**Table S2**. Influence of annealing on composition and transport properties of films. All measurements were performed at RT

| Sample | Crystalline phases | S/Ti | S(µV/K) | ρ(Ωcm) |
|---|---|---|---|---|
| **Before anneal.** | TiS$_3$ | 3.20 ±0.20 | -546 ± 20 | 0.7± 0.1 |
| **After anneal.** | TiS$_3$ | 2.52 ±0.20 | -420 ± 20 | 0.40 ±0.05 |

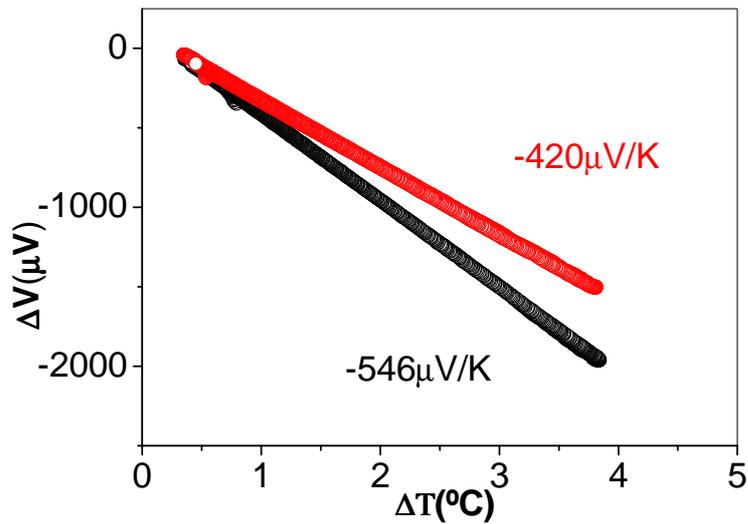

**Figure S8**: Seebeck values of films before (black dots) and after annealing at 360ºC (red dots)

### 8. Mono- and bilayer electronic band structure

In Figure S9, we report the band structure of mono- and bi-layer TiS$_3$ as obtained by using the Quantum-Espresso package, with the same physical parameters as for the band structure of the clean bulk TiS$_3$, as reported in the main text. From this calculation we see that the band-gap is similar in magnitude to the bulk (~0.9 eV vs 0.8 eV), as expected considering that the system is layered and the van der Waals interaction between the layers is weak. This is also seen by the comparison of the band structures for the mono- and bi-layer TiS$_3$: they are almost indistinguishable. In this figure, we have chosen to rigidly superimpose the valence bands to show the small differences in the conduction bands. For this calculation, to remove any spurious interaction, we have included a vacuum of 9 Angstrom in the z direction, between the replica of the mono- and bi-layer.

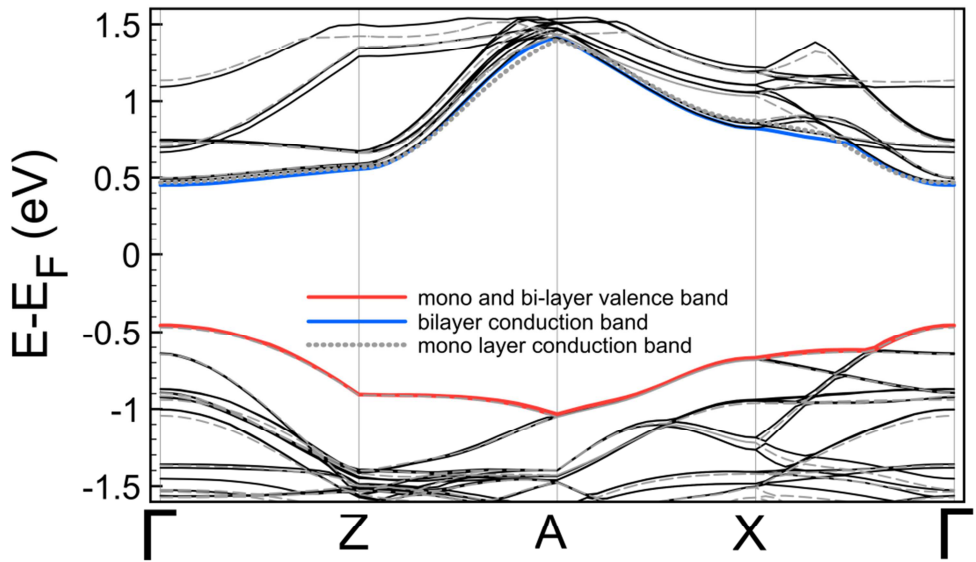

**Figure S9:** Band structures of the mono- (gray-dashed line) and bi-layer TiS3 (blue-solid line). We have rigidly shifted the bands to have the two valence bands on top of each other. The band gaps, as well as the band structures, are almost the same, an indication that the van der Waals interaction between the layers does not play a significant role in the in plane electronic structure. The path chosen is labelled according to Ref. 12.